\documentclass[ reprint, amsmath,amssymb, ajp]{revtex4-2}
\usepackage[english]{babel}
\usepackage{graphicx}
\usepackage[colorlinks=true, allcolors=blue]{hyperref}
\usepackage{braket}
\usepackage{overpic}

\begin{document}

\title{Inverse Transform Sampling for Efficient Doppler-Averaged Spectroscopy Simulations}
\author{Andrew P. Rotunno}
\affiliation{National Institute of Standards and Technology, Boulder, CO 80305, USA}
\author{Amy K. Robinson}
\affiliation{Department of Electrical Engineering, University of Colorado, Boulder, Colorado 80305, USA}
\author{Nikunjkumar Prajapati}
\author{Samuel Berweger}
\author{Matthew T. Simons}
\author{Alexandra B. Artusio-Glimpse}
\author{Christopher L. Holloway}
\thanks{christopher.holloway@nist.gov}
\affiliation{National Institute of Standards and Technology, Boulder, CO 80305, USA}
\date{\today}

\begin{abstract}
We present a thermal velocity sampling method for calculating Doppler-broadened atomic spectra, which more efficiently reaches a smooth limit than regular velocity weighted sampling.
The method uses equal-population sampling of the 1-D thermal distribution, sampling the `inverse transform' of the cumulative distribution function, and is broadly applicable to normal distributions. 
We also discuss efficiencies from eliminating velocity classes which don't significantly contribute to observed atomic lines, and comment on the application of this method in 2- and 3-dimensions. 
\end{abstract}
\maketitle

\section{Introduction}
Atomic vapors have recently grown in use as electromagnetic field sensors, utilizing spectroscopic observation of atomic states for precise, transferable measurement standards \cite{holloway2017,artusio2022modern, sedlacek2012microwave}.
A  major source of uncertainty in electromagnetically-induced transparency (EIT) spectroscopy of warm Rydberg atomic vapors comes from Doppler broadening \cite{leahy1997temperature}, where the thermal distribution of atomic velocities causes frequency shifts that result in spectral broadening.
Such effects are minimized in cold atomic systems \cite{peters2012thermometry}, or by using well-matched EIT schemes \cite{javan2002narrowing,finkelstein2019power}.
However, Doppler broadening remains a major effect in room-temperature and warmer vapor cells, broadening the D$_2$ transition in $^{133}Cs$ from its decay linewidth of $2\pi\cdot6$~MHz \cite{SteckCsData}, to a FWHM an order of magnitude larger at room temperature. 
To fit experimental spectra, we must consider this Doppler broadening, and we aim to increase efficiency when  sampling thermal velocity distribution. 


We use the illustrative case of two-photon Rydberg EIT to compute Doppler broadening, where we calculate the EIT spectrum many times over a wide sampling of velocities, and combine their spectra. 
The aim of this paper is introduce a `natural' sampling method for thermal velocity distributions, such that fewer calculations are needed to realize convergence to the Doppler-averaged curve, using a straightforward statistical transformation. 
For the same number of divisions, we find a `population' sampling method arrives at smooth transmission curves much faster than `velocity' sampling by being more dense where the resonant EIT feature occurs, while also sampling into the `wings' of the distribution.  The general method of `inverse population sampling' has been used in various contexts. Here we apply it to thermal sampling in atomic spectroscopy.

In Sec.~\ref{secAtoms}, we describe the relevant atom-photon theory, including the Maxwell-Boltzmann distribution, Doppler shifts, and atomic EIT.
In Sec.~\ref{secSample}, we describe the proposed sampling method using the inverse error function. 
In Sec.~\ref{sec:results}, we compare convergence rates between population and velocity sampling methods. 
In the appendices, we address a few special cases and extensions of the core paper. 
In Appx.~\ref{velband} we examine using a velocity cut-off for efficient line-shape calculations, in Appx.~\ref{higherDim}, we discuss the application of this method to 2D and 3D, and in Appx.~\ref{sec:lowprobe}, we confirm the new method recovers additional Doppler effects observed in Rydberg EIT. 
We review the optical master equation model in Appx.~\ref{sec:mastereq}.

\section{Background Theory}\label{secAtoms}
This section touches on relevant background theory including Doppler shifts, thermal distributions, and Doppler averaging transmission for Rydberg state EIT, with more details on transmission  given in Appx.~\ref{sec:mastereq}. 

\begin{figure}
    \centering
    \includegraphics[width=\columnwidth]{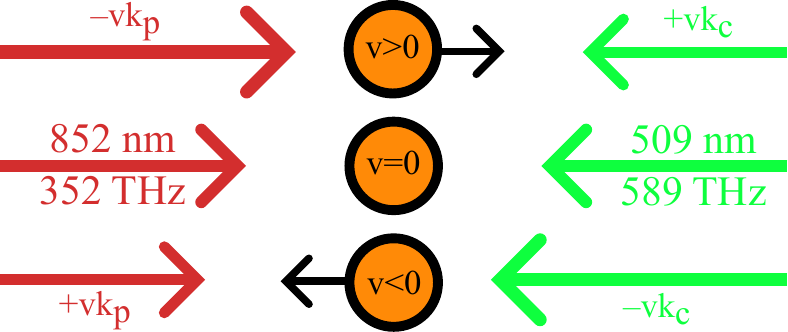}
    \caption{ Illustration of two-photon Doppler shifts for $^{133}$Cs Rydberg EIT. Photon arrow lengths illustrate observed wavelength shifts.}
    \label{fig:pedog}
\end{figure}

 \subsection{Thermal Velocities}

A population of atoms in thermal equilibrium (as in an alkali vapor cell), have 3-D velocities $\textbf{v}$ that are well-described by the Maxwell-Boltzmann distribution $f_{MB}^{3D}(\textbf{v})$.
This distribution gives the velocity density distribution as a function of temperature $T$, particle mass $m$, using the Boltzmann constant $k_B$ \cite{maxwell1860v}:
\begin{equation}
f_{MB}^{3D}(\textbf{v})dv^3 = \left(\frac{m}{2\pi{}k_BT}\right)^\frac{3}{2}\exp\left(-\frac{m|v|^2}{2k_BT}\right) dv^3\,\,\, .
\end{equation}
While across three dimensions, atoms have most likely total speed $\sqrt{\frac{2k_BT}{m}}$, we look at only one dimension, per the Doppler shift's directional selectivity.
Using $v_\sigma = \sqrt{k_BT/m}$, we have for any one dimension:
\begin{equation} \label{eq:MB1d}
f_{MB}(v) dv = \frac{1}{v_\sigma\sqrt{2\pi}}\exp\left(-\frac{v^2}{2v_\sigma^2}\right) dv \,\,\, .
\end{equation}
This distribution of velocities is shown in Fig.~\ref{fig:gausserfplots}(a).
Velocities in any one dimension are normally distributed around 0 velocity, with standard deviation $v_\sigma$.

 \subsection{Doppler Shifts}
The Doppler effect is an apparent shift in frequency of a wave when an observer or source are moving relative to one another \cite{doppler1903ueber}. 
The observed frequency shift $\Delta{}f\equiv{}f'-f_0$ from a stationary source with frequency $f_0$ is
$\Delta{}f = -\frac{v}{c} f_0  = -\frac{v}{\lambda_0}$, 
where $v$ is the speed of the observer in the wave propagation direction, $c$ the wave velocity (here, the speed of light), and wavelength $\lambda_0 = c/f_0 $.
We also write this Doppler shift as $\Delta\omega = -\textbf{v}\cdot\textbf{k} $
using angular frequency $\omega=2\pi f$, the photon's wavevector $|\textbf{k}|=2\pi/\lambda_0$, and the dot product to determine alignment with an atom's 3-D velocity $\textbf{v}$. 
The factor $v/c$ for typical atoms at room temperature is of order $10^{-6}$, but optical transitions of order $10^{14}$~Hz leave shifts at the $10^8$~Hz, nearly of order GHz, far wider than laser or decay linewidths.

\subsection{Transmission Spectrum}

In single-photon absorption, this broadening is well-characterized by a Voigt profile, the combination of Lorentzian (atomic) and Gaussian (thermal) linewidths. 
Some limits such as the `weak probe' approximation allow closed-form expressions for multi-photon EIT \cite{carr2012three,finkelstein2023practical}, so we discuss here the general case with strong probe and coupling lasers, when `brute force' parameter scan calculations are required. 
We examine a particular absorption Doppler feature of the weak probe case in Appx.~\ref{sec:lowprobe}.
In Appx.~\ref{sec:mastereq}, we detail of the calculation of two-photon spectroscopy in Rydberg EIT \cite{holloway2017,tanasittikosol2011microwave}. 
Here, we discuss the parts relevant to the thermal focus of this paper. 

We consider a three-level, two-photon system, with the `probe' laser driving the ground state to an intermediate state at 852~nm$=\lambda_p=2\pi/\textbf{k}_p$ (detuned by $\Delta_p$), and the `coupling' laser which connects the intermediate state with a Rydberg state at 509~nm$=\lambda_c=2\pi/\textbf{k}_c$ (detuned from the atomic resonance by $\Delta_c$). 
Given an atomic velocity $\textbf{v}$, the observed detuning $\Delta'$ of the probe and coupling are shifted from the lab-frame values $\Delta$: $\Delta_p'=\Delta_p-\textbf{v}\cdot{}\textbf{k}_p$ and $\Delta_c'=\Delta_c-\textbf{v}\cdot{}\textbf{k}_c$. 
When $\textbf{k}_p$ and $\textbf{k}_c$ counter-propagate, we can write the observed detunings in terms of one-dimensional velocity: $\Delta_p'=\Delta_p-v\frac{2\pi}{\lambda_p}$ and $\Delta_c'=\Delta_c+v\frac{2\pi}{\lambda_c}$.
Extension to higher dimensions is discussed in Appx.~\ref{higherDim}. 
Holding $\Delta_p=0$, any $v\neq0$ will require a shift in experimental detuning to $\Delta_c=v(\textbf{k}_p - \textbf{k}_c)$ to bring the EIT back to resonance. 
Note the vast frequency/wavelength difference in this two-photon scheme is the source of apparent detuning when atoms have $v\neq0$.
In effect, by scanning $\Delta_c$ and holding $\Delta_p=0$, we observe many velocity classes by the detuning of their EIT peak when brought into resonance. 
These constituent velocity-class transmission curves are given later in Sec.~\ref{sec:results}.
Given the large mis-match between $\lambda_p$ and $\lambda_c$, only a narrow range of velocities contribute to the Doppler-broadened EIT lineshape.


As laid out in Appx.~\ref{sec:mastereq}, the atomic coherence $\rho_{12}(v, \Delta_c)$ determines probe transmission spectrum across $\Delta_c$ for a particular $v$.
Calculation of steady-state $\rho_{12}$ is the most time-intensive step, as it must be computed multiple times over a range of $v$ in order to combine them. 
We integrate $\rho_{12}$ over velocities, weighted by the Maxwell-Boltzmann distribution, yielding the Doppler-broadened spectrum:
\begin{equation}\label{dopavg}
\rho_{12,D}(\Delta_c) = \int_{-\infty}^{\infty} \rho_{12}\left(v, \Delta_c\right)f_{MB}(v)~dv \,\,\, .
\end{equation}
We seek to minimize the number of calculations required to determine this Doppler-averaged transmission curve, which is is the primary topic of this paper.

\section{Velocity Sampling}\label{secSample}
\begin{figure}
    \centering
    \begin{overpic}[width=\columnwidth]{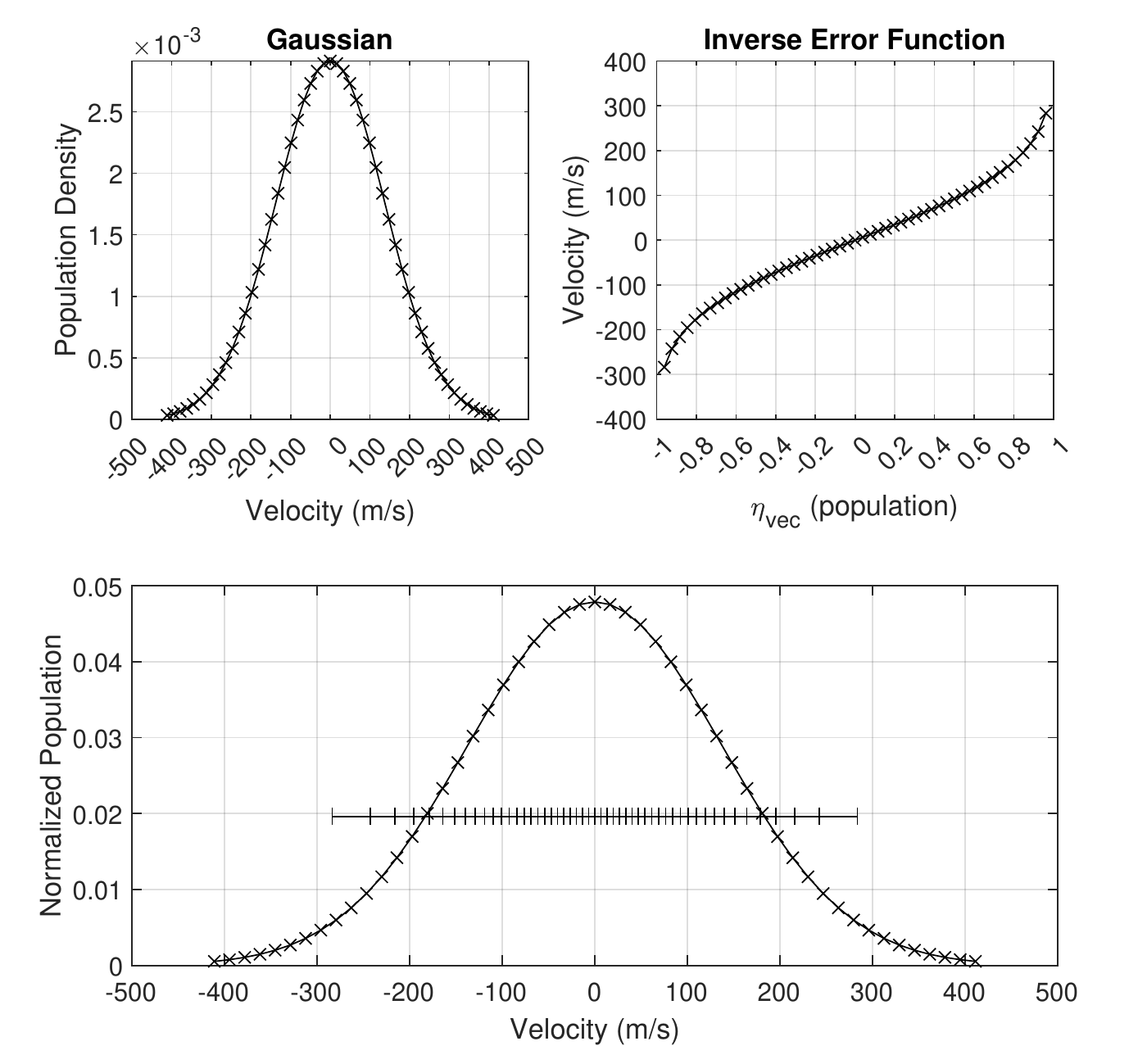}
    \put(14,80){(a)} \put(60,80){(b)}  \put(13,37){(c)}  \end{overpic}
    \caption{Contrasting sampling methods, using $N_v=51$ for $^{133}$Cs at 300~K, where $v_\sigma\approx137$~m/s. 
    (a) An equal-velocity sampling uses Gaussian statistics (3-$\sigma$, Eq.~\ref{eq:MB1d}), varying population in each sample.
    (b) An equal-population partition uses the inverse transform (Eq.~\ref{scaledef}), varying velocity density. 
    (c) Plotting both on the same probability scale, showing the dynamic velocity spacing of the population sampling method.}
    \label{fig:gausserfplots}
\end{figure}

The most straightforward approach 
to evaluate the velocity-space integral of Eq.~\ref{dopavg} is by constructing a discrete equally-spaced `scan' of velocities over a few standard deviations, to cover nearly all of the atomic ensemble's exhibited velocities. 
We refer to this as the ``velocity sampling'' method, and we plot a representative sampling in Fig.~\ref{fig:gausserfplots}(a). 
Each bin of velocity is weighted by the Maxwell-Boltzmann distribution to determine what fraction of the population is between $v$ and $v+dv$.
This renders Eq.~\ref{dopavg} discretized for computation as:
\begin{equation}\label{dopavgsigmas}
\rho_{12,D}(\Delta_c) = \frac{\Delta v_i}{v_\sigma\sqrt{2\pi}} \sum_{v_i=-3v_\sigma}^{3v_\sigma} \rho_{12}\left(v_i,\Delta_c\right)\exp\left(-\frac{v_i^2}{2v_\sigma^2}\right) 
\end{equation} 
where common factors are brought outside the sum. 
In Eq.~\ref{dopavgsigmas}, the sample $v_i$ spans three standard deviations, with equal spacing $\Delta{}v_i$. 
For a symmetric summation over $N_s$-$\sigma$ (where $N_s=3$ in the above example) using $N_v$ points, we have velocity resolution $\Delta{}v_i = 2N_sv_\sigma/N_v$ across the entire range. 
An even velocity sampling means that in order to sample densely near $v_i=0$, one must also densely sample the distribution's `wings,' where the contribution to a resonant interaction is minimal, both by being off-resonance and diminished by the Maxwell-Boltzmann density. 

\subsection{Local Refinement}
To significant effect in computational efficiency, we can break the summation of Eq.~\ref{dopavgsigmas} into parts: one with `fine' velocity spacing at the region of interest, the rest remains `coarse.' 
This fine segment can be centered around arbitrary velocities ($v\neq0$), in the event that detuning in either field requires an off-resonant calculation. 
Imposing bounds at $v_\downarrow$ and $v_\uparrow$, and using `fine' and `coarse' velocity spacing $\Delta{}v_f\ll\Delta{}v_c$, we can split Eq.~\ref{dopavgsigmas} into partial sums with different sample spacing and bounds:
\begin{equation}\label{dopavgsigmasum}
\begin{split}
\rho_{12,D}(\Delta_c) =& \\ 
~\frac{\Delta v_c}{v_\sigma\sqrt{2\pi}}&
\sum_{v_i=v_\uparrow}^{3v_\sigma} \rho_{12}\left(v_i,\Delta_c\right)e^{\frac{-v_i^2}{2v_\sigma^2}} \\
+\frac{\Delta v_f}{v_\sigma\sqrt{2\pi}}& \sum_{v_i=v_\downarrow}^{v_\uparrow} \rho_{12}\left(v_i,\Delta_c\right)e^{\frac{-v_i^2}{2v_\sigma^2}} \\
+\frac{\Delta v_c}{v_\sigma\sqrt{2\pi}} &
 \sum_{v_i=-3v_\sigma}^{v_\downarrow}
 \rho_{12}\left(v_i,\Delta_c\right)e^{\frac{-v_i^2}{2v_\sigma^2}}
\end{split} \,\,\,\,\, .
\end{equation} 

In practice, the bounds $v_\uparrow$ and $v_\downarrow$ are brought in until their cutoff is observed in the `wings' of the calculated EIT peak, typically a fraction of $1\sigma$, and the coarse velocity spacing accounts for the rest of the overall absorption.
We must re-weight these for combination by  $\Delta{}v_f$ and $\Delta{}v_c$, as velocity bin size scales the Maxwell-Boltzmann density.
This splitting into coarse and fine velocity ranges allows one to focus computational resources on near-resonant velocity classes, which heavily depend on small changes in $v$, while economizing on far off-resonant velocity classes for which the absorption profile is nearly flat.

\subsection{Population Sampling Method}
We are motivated by this \emph{ad hoc} sampling in the region of interest to construct a `natural' velocity sampling that over-samples near resonance, and under-samples in the wings, for computational efficiency in the resonant case. 
To this end, we sample velocities that are equally spaced in the Gaussian-distributed population or percentile space, such that each step represents an equal population of atoms, $1/N_v$ for $N_v$ velocity bins. 
In Fig.~\ref{fig:gausserfplots}(b,c), we show this equal-population method's dynamic velocity sampling, using $N_v=51$ for illustration.
This method of percentile sampling is broadly applicable, often given the name ``inverse transform sampling'' \cite{olver2013fast}, or probit analysis \cite{finney1952probit} when applied to population statistics.

We begin to calculate the velocity of each partition of population using the error function $\textrm{erf}(v)$ \cite{abramowitz1964handbook}.
A Gaussian's cumulative distribution function is $\frac{1}{2}\left(1+\textrm{erf}\left(\frac{v-\mu}{\sqrt{2}\sigma}\right)\right)$, for mean $\mu$ and standard deviation $\sigma$. 
The error function spans $v=(-\infty,\infty)$, and yields a value which integrates signed population away from the mean, from $0$ to $v$:
\begin{equation}\label{erfdef}
\textrm{erf}(v) \equiv \frac{2}{\sqrt{\pi}}\int_0^ve^{-t^2}dt    
\end{equation}

To perform the inverse transform sampling, we use the inverse of the error function $\textrm{erf}^{-1}(\eta)$ or $\textrm{inverf}(\eta)$, where $\eta$ spans the signed population space $(-1,1)$ with $N_v$ points.
The inverse error function transforms a linearly-spaced (or randomized) population into an inhomogeneously-spaced distribution of velocity values.
The function  $\textrm{erf}^{-1}(\eta)$ does not have a closed form expression, and is implemented numerically \cite{wichura1988algorithm}.
 
In pseudo-code, the velocity distribution vector $v_{vec}$ of a Gaussian-distributed, normalized population is:
\begin{equation}\label{scaledef}
     v_{vec} = v_\sigma\sqrt{2}\cdot{}\textrm{erf}^{-1}(\eta_{vec})
\end{equation}
where $\eta_{vec}$ is the discretized equal-spaced population sampling vector from $-1$ to $1$ with $N_v$ elements. 
We plot $v_{vec}$ against $\eta_{vec}$ in Fig.~\ref{fig:gausserfplots}(b). 
The endpoints $-1,1$ yield infinities, so the full vector $\eta_{vec}$ is generated with $N_v+2$ points, then we use elements $[2]$ to $[end-1]$, for starting index~1. 
If $N_v$ is odd, the center point 0 is included. 
In the limit of high $N_v\gtrsim100$, the gap $dv$ near 0 approaches $v_\sigma\cdot\sqrt{2/\pi}/N_v$. 
Since $\sqrt{2/\pi}\approx0.8$, this sampling is approximately as dense as equal-spaced velocities spanning only a symmetric $0.4\sigma$, while actually sampling asymptotically into the distribution wings. 
While this inverse-transform sampling method works well for resonant EIT, it is less efficient  when significant detuning is employed, whereas the `coarse/fine' method works well for arbitrary $v$.

The calculation includes a new normalization factor, dividing by $N_v$, which we can do with two different heuristic interpretations:
First, we can consider that each velocity class is only a portion ($1/N_v$) of the total atom number. 
In this case, we \textit{sum} over the velocities to get total susceptibility from all of them.
In a second interpretation, each $\rho_{12}$ represents the transmission spectrum if the entire atomic population had that velocity class. 
In this case, we \textit{average} over the velocities, to see how the total population transmits light. 
Note in either case, the sum of $\rho_{12}$ or susceptibility over velocities is divided by $N_v$, and each are equally weighed. 

Having combined susceptibilities, we can compute total transmission using Beer's law, Eq.~\ref{beer} in Appx.~\ref{sec:mastereq}.
We note that adding susceptibilities (inside the exponential) and taking one exponential is nominally fewer computations than taking an exponential of each susceptibility and multiplying those transmission values, as the two are mathematically equivalent. 
Additionally, this method skips the Maxwell-Boltzmann weighting multiplication, which also grows with $\mathcal{O}(N_v)$, although these operations are insignificant with respect to finding the steady-state atomic coherence matrix $\rho$, which remains the largest computational load. 




\section{Results}\label{sec:results}
\begin{figure}
    \centering
    \includegraphics[width=\columnwidth]{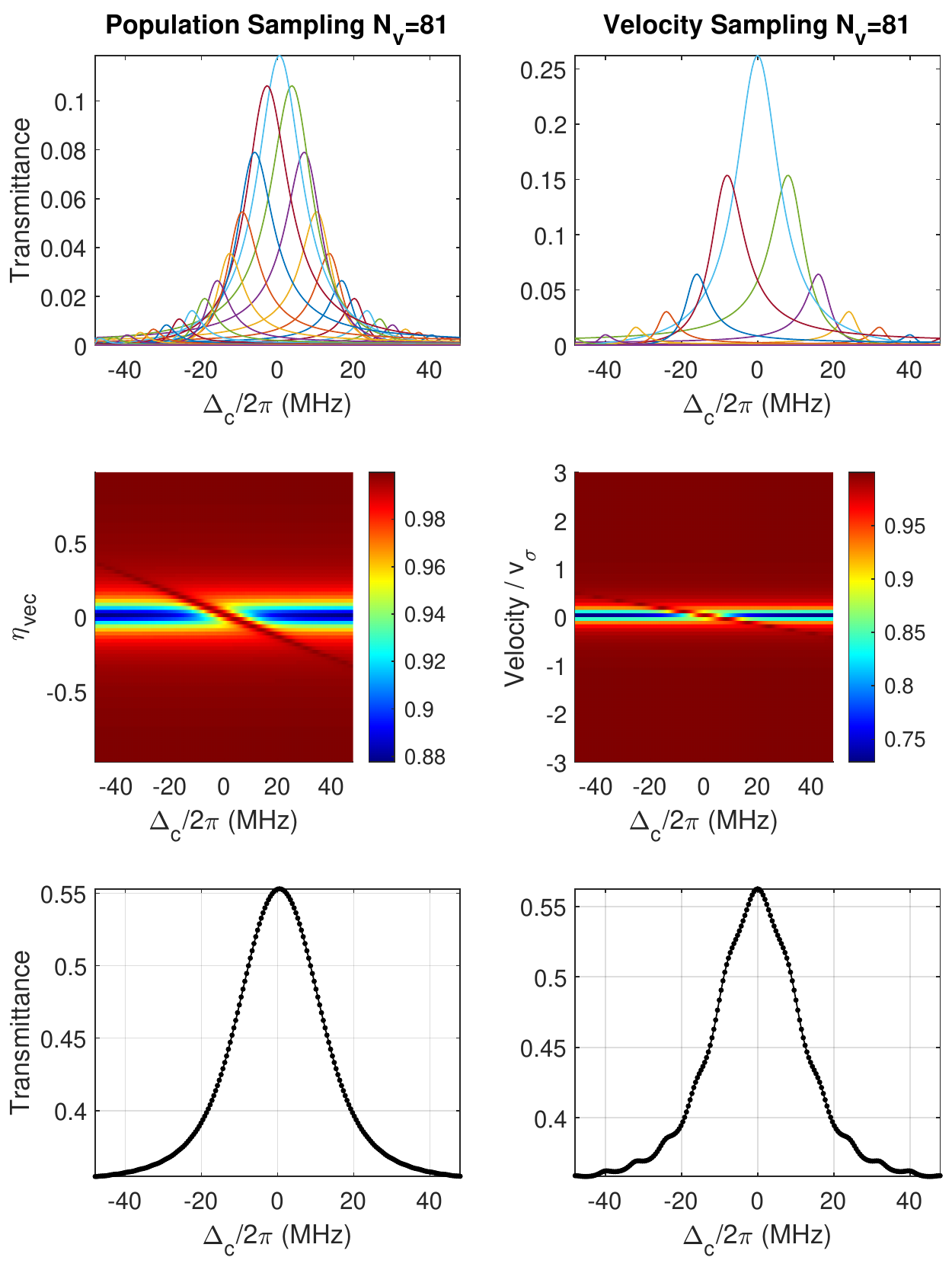}
    \caption{Comparison of simulation results using population (left) vs. velocity (right) sampling methods, using parameters listed in Sec.~\ref{sec:results}. 
    Top row: Transmission spectra (offset) over detuning for many constituent velocity samples. 
    Middle row: Transmission for a two-dimensional parameter scan over detuning and velocity.
    Bottom row: Doppler-averaged total transmittance spectra. }
    \label{fig:surfSampling}
\end{figure}
\begin{figure*}
\centering
    \includegraphics[width=\textwidth]{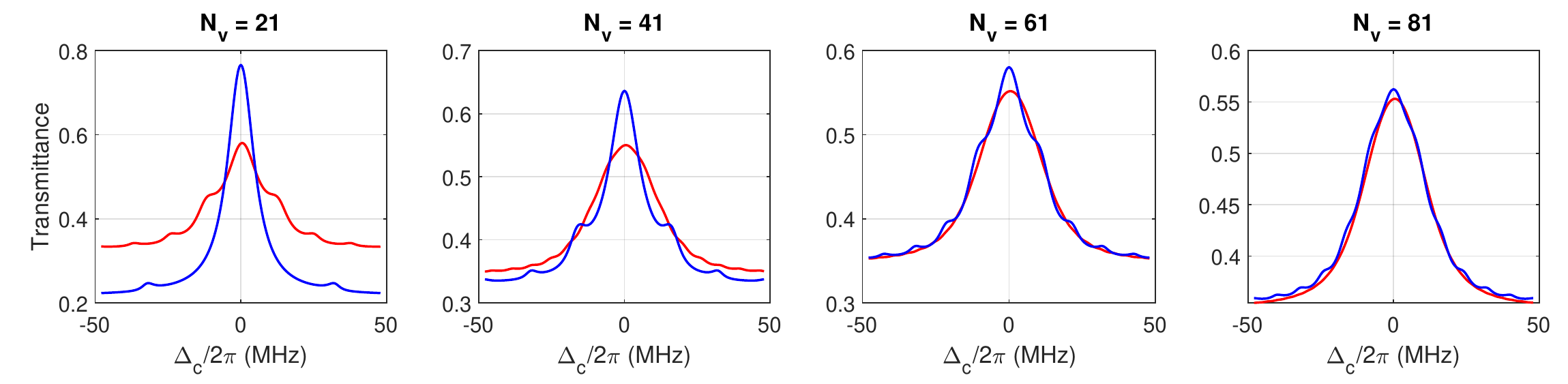}
    \caption{ Comparison of calculated transmission curves over coupling detuning for the population sampling method (red) and the velocity sampling method (blue), as $N_v$ is varied (plot titles). }
    \label{fig:nvelCompare}
\end{figure*}
We illustrate the new ``population'' sampling method with the ``velocity'' sampling method in Fig.~\ref{fig:surfSampling}. 
We compare computed EIT transmission spectra of Cesium's 55$S_{1/2}$ state, holding the same number of sampling points ($N_v=81, N_\Delta=201$) and all other parameters held constant ($\lambda_p\approx852.3$~nm, $\lambda_c\approx509.4$~nm, $\Omega_p/2\pi\approx18$~MHz, $\Omega_c/2\pi\approx3.4$~MHz, $\Delta_p=\Delta_c=0$, temperature of 293~K, cell length 75~mm), using Eq.~\ref{scaledef} to generate $v_{vec}$ for population sampling on the left, and velocity sampling over 3$\sigma$ on the right side.
This comparison shows that population sampling converges to a smooth transmission curve with the same parameters, whereas velocity sampling exhibits sampling artifacts. 

The first row in Fig.~\ref{fig:surfSampling} plots individual velocity sample transmission curves (i.e., the EIT signal), where we have offset the transmission of each curve to illustrate how they combine into the total transmission curve.
The middle row of Fig.~\ref{fig:surfSampling} represents the same information in a surface plot over detuning horizontally, and velocity sample point vertically, using the color to represent transmission value. 
The bottom plot of Fig.~\ref{fig:surfSampling} shows the resultant curve when combining velocities for the two cases.
The population sampling method provides denser sampling in the resonant region for equal $N_v$, while `bumpy' artifacts remain in velocity sampling case, due to insufficient sampling density. 

The primary objective of this effort is to reduce the number times we have to calculate the steady-state solution for $\rho_{12}$ before reaching a `smooth' simulated spectroscopy absorption profile.
We show a qualitative comparison of transmission curves for a range of $N_v$ in Fig.~\ref{fig:nvelCompare}. 
The low density velocity sampling is apparent in both cases for low $N_v$, but the population method is seen to converge to a smooth curve for moderate values of $N_v$.

\begin{figure}
    \centering
    \includegraphics[width=\columnwidth]{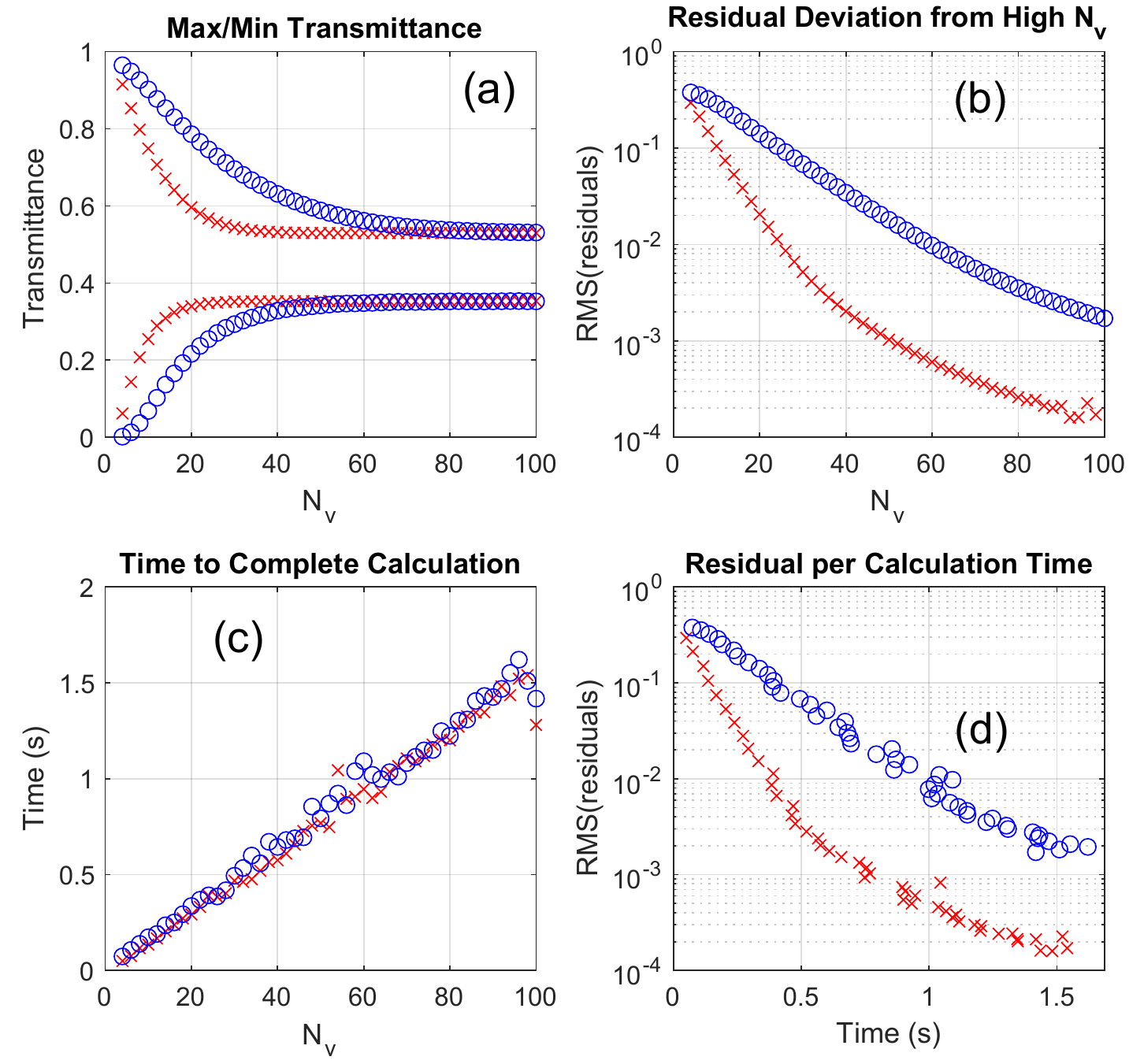}
    \caption{Calculation time using odd $N_v$ for the population method in red \textcolor{red}{$\times$}'s and the velocity method in blue \textcolor{blue}{$\circ$}'s.
    (a) Total transmission maxima and minima. 
    (b) RMS of differences between each spectrum and the population curve at $N_v=101$. 
    (c) Time per spectrum calculation. 
    (d) Residual RMS against calculation time. 
    }
    \label{fig:convergence}
\end{figure}

In Fig.~\ref{fig:convergence}, we quantitatively compare results from both methods while varying $N_v$ to see when each method converges to a smooth transmission curve. 
To measure the computational convergence rates as $N_v$ is increased, we use two metrics: the extrema transmission values of the spectrum in Fig.~\ref{fig:convergence}(a), and the `error' quantified as RMS of residuals from the population method at $N_v=101$ in Fig.~\ref{fig:convergence}(b). 
The actual values of $N_v$ required for convergence will depend heavily on other parameters in any particular application. 
Plotted in Fig.~\ref{fig:convergence}(c), we show computation time required between the two methods, which is nearly equal, dominated by the steady-state $\rho$ calculation, which occurs $N_v\times N_\Delta$ times in either method.
The highest $N_v$ were calculated first, and ran slightly faster than the linear trend of $\mathcal{O}(N_v)$. 
The advantage in computation speedup is illustrated in Fig.~\ref{fig:convergence}(d), where we plot the RMS residual `error' against computation time. 
The population sampling method reaches 1\% RMS residuals in less than half the computation time as the velocity sampling method.
By over-sampling velocities near zero, far fewer samples are needed to obtain accurate spectra.

\section{Conclusion}\label{sec:conc}
We have presented a method for calculating Doppler-broadened transmission spectra for Rydberg EIT measurements with more efficient sampling over population than velocity. 
Rather than using a Maxwell-Boltzmann weighting over an equal-spaced velocity distribution, we use an equal-population partition to `naturally' sample densely near zero velocity, and sparsely sample into the thermal distribution's wings.
This significantly lowers the required number of velocity-sampling curves to be calculated before `smooth' convergence of the transmission curves is obtained.
In this work, we reach 1\% error in less than half the time of the typical method for the chosen parameters. 
Conversion of Doppler-sampling scripts to the new method is as simple as adjusting the velocity vector that is sampled over, and changing the Maxwell-Boltzmann weighting operation of Eq.~\ref{dopavg} into an average over velocities. 


\appendix

\section{Cheap (Un-physical) Calculations} \label{velband}
\begin{figure}
    \centering
    \includegraphics[width=\columnwidth]{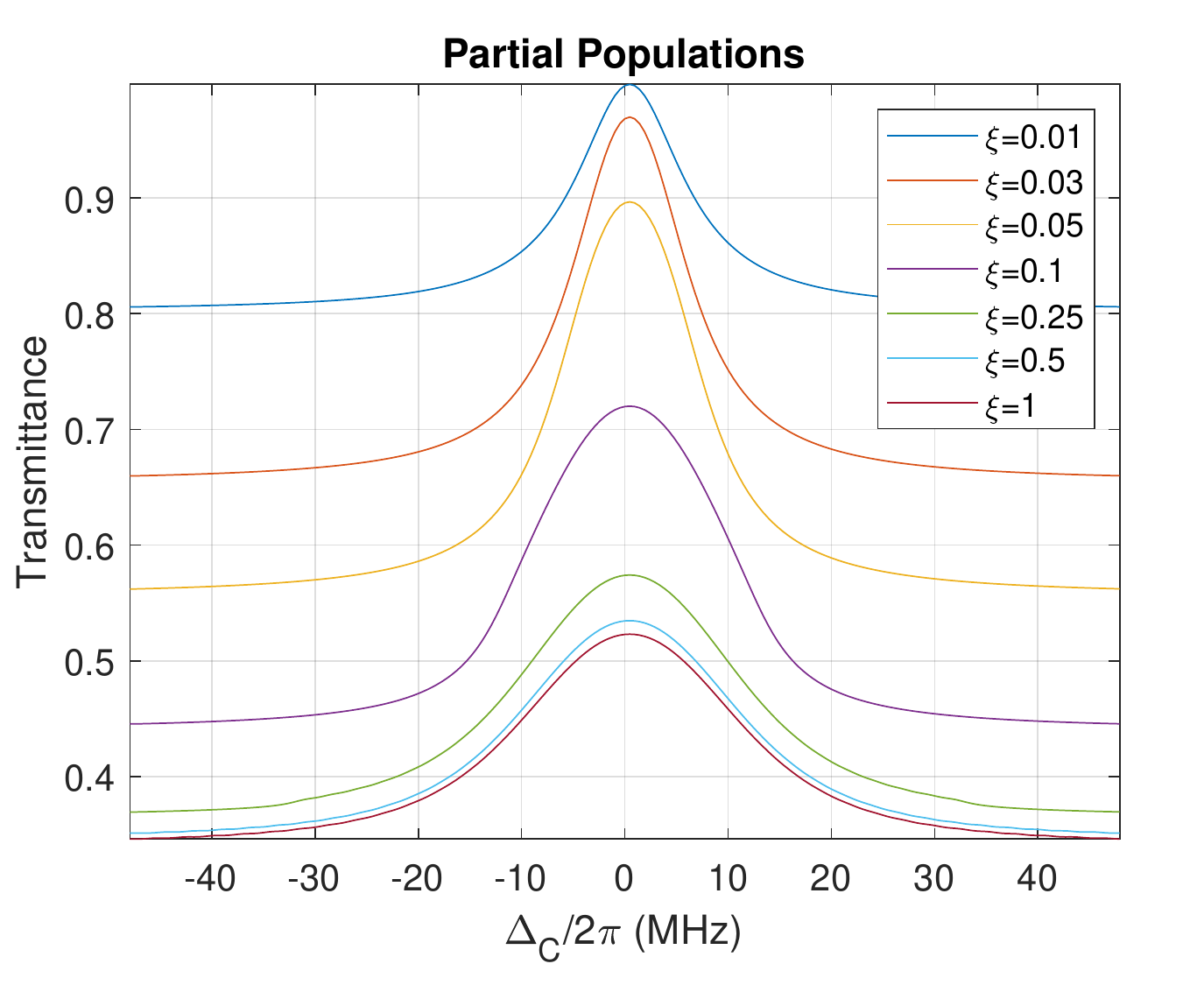}
    \caption{Transmission spectra from partial velocity populations $0<\xi<1$ around $v=0$. }
    \label{fig:velband}
\end{figure}
There is one further simplification that can be implemented for rough line-shape calculations for the resonant case (including for our typical application of Autler-townes splitting \cite{holloway2017}), but does not give physical total transmission values. 
Looking at the velocity dependence in the middle row of Fig.~\ref{fig:surfSampling}, we notice that only the central portion near $v=0$ experiences resonant EIT, and the rest is absorbed independent of $\Delta_c$. 
Instead of calculating the entire velocity distribution, we can restrict the bounds of $v_{vec}$ to just the lowest velocities, which contribute most to resonant spectral features. 
This is synonymous with taking the above `fine' partition alone, and tossing the `coarse' parts. 
Implementing this is as easy as using a tighter bound, $0<\xi<1$ to generate $\eta_{vec}$ from $-\xi$ to $\xi$.
Maintaining sampling density now requires smaller $N_v$ by a factor of $\xi$, reducing computation time. 
This sub-sampling reveals the approximate line shape of EIT features, without giving an appropriate total transmission, lacking the remaining probe absorption.

We illustrate the effect of varying $\xi$ in Fig.~\ref{fig:velband}, where we adjust $N_v$ to maintain sampling density, such that the computation time lowers by a nominal factor of $\xi$ while giving approximate line shapes. 
As a result of this `velocity band' narrowing, we use a scaling factor of $N_0 \rightarrow \xi{}N_0 / N_v$ in Eq.~\ref{chi1}, to account for the sub-sampling of the population space. 
One can lower $\xi$  until it `cuts into the wings' of the EIT feature, which is more evident for smaller probe powers than the one in our simulations, which broadens the line to order $\Omega_p\approx$18~MHz.
Lowering $\xi$ crucially allows us to reduce $N_v$ at the same time to achieve the same sampling density. 
This method is most useful for resonant, poorly Doppler-matched beams, where only the near-zero velocities exhibit EIT. 

\section{Higher Dimension Extension}\label{higherDim}
This sampling method can be easily extended to apply to EIT configurations with three or more lasers which are non counter-propagating, as in planar (2D) three-photon configuration \cite{grynberg1977doppler, thaicharoen2019electromagnetically, carr2012three}, or in general 3D. 
These schemes use angle-tuning to better remove residual Doppler mis-match, depending on the choice of wavelengths used \cite{shaffer2018read}. 

The Gaussian velocity distribution applies independently over each direction. 
Therefore in 2-D, we can loop over both velocity directions (i.e. for each $v_x$ calculate all $v_y$ combinations), sampling the same $v_{vec}$ values, independently. 
For each combination of $v_x$ and $v_y$, we then calculate the Doppler-adjusted detuning $\Delta'=\Delta - \textbf{v}\cdot{}\textbf{k}$ for each laser, where for convenience, at least one laser can be along $\hat{x}$, whose effective detuning is only changed in the outer loop. 
An off-axis beam's Doppler adjustment $\textbf{v}\cdot{}\textbf{k}$ can be inferred from the atom's trajectory $\theta = \tan^{-1}\left(\frac{v_y}{v_x}\right)$ and total speed $|v| = \sqrt{v_x^2+v_y^2}$.
An extra normalization factor of $N_0 / N_v^2$ must be included before the sum over both dimensions, as each sample represents a further division of the population space from the 1-D case. 
This approach is easily extended from two to three dimensions, nominally increasing calculation time from $\mathcal{O}(N_v^2)$ to  $\mathcal{O}(N_v^3)$ from 2-D to 3-D respectively, making reductions in $N_v$ all the more significant.

\section{Low-Probe Case}\label{sec:lowprobe}
\begin{figure}
    \centering
    \includegraphics[width=\columnwidth]{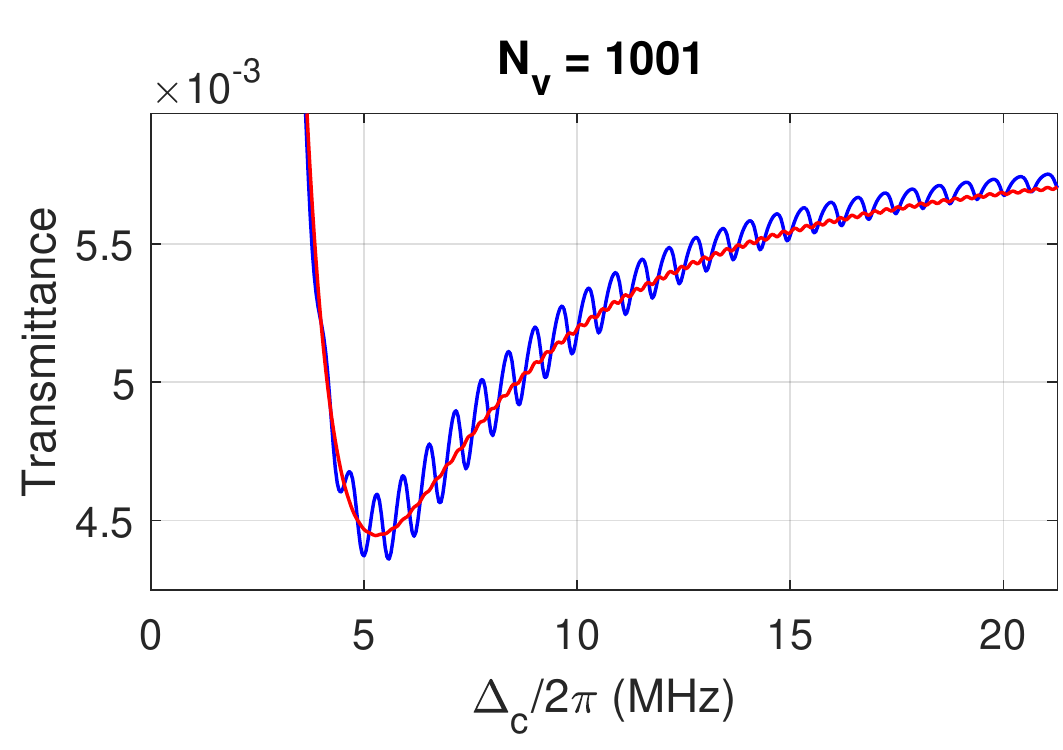}
    \caption{Doppler-based absorption enhancement for a small probe power, comparing results from either sampling method.
    }
    \label{fig:lowprobe}
\end{figure}

In addition to Doppler broadening of lines, velocity-dependent absorption curves give other Doppler effects, such as enhanced \emph{absorption} on either side of an EIT line \cite{carr2012three}. 
To ensure our method captures these features as well, we compare the population and velocity sampling methods in Fig.~\ref{fig:lowprobe}.
This calculation uses $\Omega_p/2\pi\approx0.40$~MHz, and $\Omega_c/2\pi\approx3.1$~MHz. 
Since this feature is near the resonance, it requires significantly higher sampling density than cases with significant probe Rabi broadening, as in the rest of this paper. 
Again, the population sampling method appears to converge closer to a smooth curve for the same $N_v$ compared to the velocity sampling method.

\section{Master-equation model}\label{sec:mastereq}
We find the time-independent solutions of a master-equation model for calculating EIT signals. 
The model presented here is for $^{133}$Cs atoms, although a similar model for $^{87}$Rb is presented in Ref. \cite{holloway2017}.  
The power of the probe beam measured on the detector (the EIT signal, i.e., the probe transmission through the vapor cell) is given by \cite{yarin}
\begin{equation}
P_{out}=P_{in} \exp\left(-\frac{2\pi L \,\,{\rm Im}\left[\chi\right]}{\lambda_p}\right)=P_{in} \exp\left(-\alpha L\right) \,\,\, ,
\label{beer}
\end{equation}
where $P_{in}$ is the power of the probe beam at the input of the cell, $L$ is the length of the cell, $\lambda_p$ is the wavelength of the probe laser,  $\chi$ is the susceptibility of the medium seen by the probe laser, and $\alpha=2\pi{\rm Im}\left[\chi\right]/\lambda_p$ is Beer's absorption coefficient for the probe laser.  The susceptibility for the probe laser is related to the density matrix component ($\rho_{21}$)  by the following \cite{berman}
\begin{equation}\label{chi1}
\chi=\frac{2\,{\cal{N}}_0\wp_{12}}{E_p\epsilon_0} \rho_{21_D} =\frac{2\,{\cal{N}}_0}{\epsilon_0\hbar}\frac{(d\, e\, a_0)^2}{\Omega_p} \rho_{21_D}\,\,\, ,
\end{equation}
where $d=2.02$ is the normalized transition-dipole moment \cite{SteckCsData} for the probe laser and $\Omega_p$ is the Rabi frequency for the probe laser in units of rad/s. 
The subscript $D$ on $\rho_{21}$ presents a Doppler averaged value, per this paper's theme. 
${\cal{N}}_0$ is the total density of atoms in the cell and is given by
\begin{equation}
{\cal{N}}_0= \frac{p}{k_B T} \,\, ,
\label{nn}
\end{equation}
where $k_B$ is the Boltzmann constant, $T$ is temperature in Kelvin, and the pressure $p$ (in units of Pa) is given by \cite{SteckCsData}
\begin{equation}
p = 10^{-217.3571+\frac{1088.676}{T}-0.08336185\cdot{}T+94.88752\log_{10}(T)}
\label{ppp}
\end{equation}
In Eq.~(\ref{chi1}), $\wp_{12}$ is the transition-dipole moment for the $\ket{1}$-$\ket{2}$ transition, $\epsilon_0$ is the vacuum permittivity, and $E_p$ is the amplitude of the probe laser E-field.

The density matrix component ($\rho_{21}$) is obtained from the master equation \cite{berman}
\begin{equation}
\dot{\boldsymbol{\rho}}=\frac{\partial \boldsymbol{\rho}}{\partial t}=-\frac{i}{\hbar}\left[\mathbf{H},\boldsymbol{\rho}\right]+\boldsymbol{\cal{L}} \,\,\, ,
\label{me}
\end{equation}
where $\mathbf{H}$ is the Hamiltonian of the atomic system under consideration and ${\boldsymbol{\cal{L}}}$ is the Lindblad operator that accounts for the decay processes in the atom. The $\mathbf{H}$ and $\boldsymbol{\cal{L}}$ matrices for the three different tuning schemes are given below.

We numerically solve these equations to find the steady-state solution for $\rho_{21}$ for various values of Rabi frequency ($\Omega_i$) and detunings ($\Delta_i$). This is done by forming a matrix with the system of equations for $\dot{\rho}_{ij}=0$. The null-space of the resulting system matrix is the steady-state solution.  The steady-state solution for $\rho_{21}$ is then Doppler averaged~\cite{berman}
\begin{equation}
\rho_{21_D}=\frac{1}{\sqrt{\pi}\,\, v_{\sigma}}\int_{-3v_{\sigma}}^{3v_{\sigma}}\rho_{21}\left(\Delta'_p,\Delta'_c\right)\,\,e^{\frac{-v^2}{v_{\sigma}^2}}\,\,dv\,\,\, ,
\label{doppler}
\end{equation}
where $v_{\sigma}=\sqrt{k_B T/m}$ and $m$ is the mass of the atom. We use the case where the probe and coupling laser are counter-propagating. Thus, the frequency seen by the atom moving toward the probe beam is upshifted by $2\pi v/\lambda_p$ (where $v$ is the velocity of the atoms), while the frequency of the coupling beam seen by the same atom is downshifted by $2\pi v/\lambda_c$.  The probe and coupling beam detuning is modified by the following
\begin{equation}
\Delta'_p=\Delta_p-\frac{2\pi}{\lambda_p}v \,\,\,{\rm and}\,\,\,
\Delta'_c=\Delta_c+\frac{2\pi}{\lambda_c}v \,\,\, .
\label{doppler2}
\end{equation}

For the three level system, the Hamiltonian can be expressed as:
\begin{equation}
\begin{footnotesize}
H=\frac{\hbar}{2}\left[\begin{array}{cccc}
0 & \Omega_p & 0\\
\Omega_p & -2\Delta_p' & \Omega_c \\
0 & \Omega_c & -2(\Delta_p'+\Delta_c') \\
\end{array}
\right]\,\, ,
\end{footnotesize}
\label{H4}
\end{equation}
where $\Delta_p'$ and $\Delta_c'$ are the Doppler-shifted detunings, and $\Omega_p$ and$\Omega_c$ are the Rabi frequencies associated with the probe laser and 
coupling laser respectively. 
The detuning for each field is defined as
\begin{equation}
\Delta_{p,c}=\omega_{p,c}-\omega_{o_{p,c}} \,\,\, ,
\label{rabi}
\end{equation}
where $\omega_{p,c}$ are the on-resonance angular frequencies of transitions $\ket{1}$-$\ket{2}$, $\ket{2}$-$\ket{3}$, and $\ket{3}$-$\ket{4}$, respectively;
and $\omega_{p,c}$ are the angular frequencies of the probe laser and coupling laser, respectively.
The Rabi frequencies are defined as $\Omega_{p,c}=|E_{p,c}|\frac{\wp_{p,c}}{\hbar}$,
where $|E_{p, c}|$ are the magnitudes of the E-field of the probe laser, the coupling laser, and the RF source, respectively.
Finally, $\wp_p$ and $\wp_c$ are the atomic dipole moments corresponding to the probe, coupling, and RF transitions.

For the three-level system, the ${\cal{L}}$ matrix is given by
\begin{equation}
\footnotesize
{\cal{L}}=\left[\begin{array}{cccc}
\Gamma_2 \rho_{22} & -\gamma_{12}\rho_{12} & -\gamma_{13}\rho_{13} \\
-\gamma_{21}\rho_{21} & \Gamma_3 \rho_{33}-\Gamma_2 \rho_{22} & -\gamma_{23}\rho_{23}\\
-\gamma_{31}\rho_{31} & -\gamma_{32}\rho_{32} & -\Gamma_3 \rho_{33} 
\end{array}
\right] \,\,\, ,
\label{H4}
\end{equation}
where $\Gamma_{i}$ are the state-to-state spontaneous decay rates, and $\gamma_{ij}=(\Gamma_i+\Gamma_j)/2$ use the total decoherence rates of either state.
We have ground state $\Gamma_1~=~0$, D$_2$ line \cite{SteckCsData}  $\Gamma_2=2\pi\cdot~5.222$~MHz, and state-to-state decay from $\ket{55S_{1/2}}\rightarrow\ket{6P_{3/2}}$ of  $\Gamma_3=2\pi\cdot~308$~Hz.
The decoherence terms $\gamma_{3,i}$ use total Rydberg decay rate  at $T=300$~K of $2\pi\cdot~83.867$~kHz \cite{vsibalic2017arc}. 
\bibliography{main}

\end{document}